\documentclass[letterpaper, 10pt, conference]{ieeeconf}

\IEEEoverridecommandlockouts

\newtheorem{theorem}{Theorem}

\newtheorem{lemma}{Lemma}
\newtheorem{remark}{Remark}
\newtheorem{definition}{Definition}
\newtheorem{proposition}{Proposition}

\newtheorem{assumption}{Assumption}
\usepackage{physics}
\usepackage[dvipsnames]{xcolor}
\usepackage{tikz}
\usepackage[none]{hyphenat}
\usetikzlibrary{external}
\DeclareMathOperator*{\argmin}{argmin}
\DeclareMathOperator*{\argmax}{argmax}
\usetikzlibrary{mindmap,trees}
\usetikzlibrary{arrows}
\usetikzlibrary{calc}
\usepackage{amsmath,amsfonts,amssymb,color}
\usepackage{epsfig}
\usepackage{psfrag}
\usepackage{algorithm}
\usepackage{algpseudocode}
\usepackage{array}
\usepackage{epstopdf}
\usepackage{tikz, subcaption, cite}
\usepackage{comment}
\renewcommand{\theenumi}{\roman{enumi}}%
\newcommand\scalemath[2]{\scalebox{#1}{\mbox{\ensuremath{\displaystyle #2}}}}
\usepackage{mathtools}

\DeclarePairedDelimiter{\diagfences}{(}{)}
\newcommand{\diag}{\operatorname{diag}\diagfences}

\title{\LARGE \bf The Effect of Behavioral Probability Weighting in a Simultaneous Multi-Target Attacker-Defender Game}
\author{Mustafa Abdallah,~Timothy Cason,~Saurabh Bagchi, and~Shreyas Sundaram
\thanks{This research was supported by grant CNS-1718637 from the National Science Foundation. Mustafa Abdallah, Saurabh Bagchi, and Shreyas Sundaram are with the School of Electrical and Computer Engineering at Purdue University. Email: {\tt \{abdalla0,sbagchi,sundara2\}@purdue.edu}. Timothy Cason is with the Krannert School of Management at Purdue University. Email: {\tt{cason@purdue.edu}}.}.%
}

\begin{document}
\sloppy

\maketitle
\thispagestyle{empty}
\pagestyle{empty}

\begin{abstract}
We consider a security game in a setting consisting of two players (an attacker and a defender), each with a given budget to allocate towards attack and defense, respectively, of a set of nodes.  Each node has a certain value to the attacker and the defender, along with a probability of being successfully compromised, which is a function of the investments in that node by both players.  For such games, we characterize the optimal investment strategies by the players at the (unique) Nash Equilibrium. We then investigate the impacts of behavioral probability weighting on the investment strategies; such probability weighting, where humans overweight low probabilities and underweight high probabilities, has been identified by behavioral economists to be a common feature of human decision-making. We show via numerical experiments that behavioral decision-making by the defender causes the Nash Equilibrium investments in each node to change (where the defender overinvests in the high-value nodes and underinvests in the low-value nodes).
\end{abstract}


\section{Introduction}
Today's cyber-physical systems (CPS) are increasingly facing attacks by sophisticated adversaries, who are able to evaluate the susceptibility of different goal targets in the system and strategically allocate their efforts to compromise the security of the CPS~\cite{humayed2017cyber,he2016cyber}. 
In response to such intelligent adversaries, the operators (or defenders) of CPS need to allocate their (limited) security budget across many assets to best mitigate their vulnerabilities. This motivates the need to capture such interactions between attackers and defenders
and study their effects on system security. In this context, significant research has been conducted for understanding how to better secure these systems, with game-theoretical models receiving increasing attention due to their power in capturing the interactions of players (strategic attackers and defenders) in different settings~\cite{laszka2015survey,alpcan2010network,guan2017modeling,an2016stackelberg}. 

A particular class of simultaneous move games involving attackers and defenders (where the players have to choose their strategies at the same time, without first observing what the other player has done) have been studied in various contexts.  For example, the Colonel Blotto game \cite {roberson2006colonel} is a useful framework to model the allocation of a given amount of resources on different potential targets (i.e., battlefields) between the attacker and the defender. Specifically, \cite{7943422} proposed a solution of the heterogeneous Colonel Blotto game with asymmetric players (i.e., with different resources) and with a number of battlefields that can have different values. While Colonel Blotto games typically involve deterministic success functions (where the player with the higher investment on a node wins that node), other work has studied cases where the win probability for each player is a probabilistic (and continuous) function of the investments by each player~\cite{guan2017modeling}. 

In these works, following classical game-theoretical models of human decision-making, defenders and attackers are considered to be fully rational decision-makers who choose their actions to  maximize their expected utilities. However, a seminal model called {\it prospect theory} (introduced by Kahneman and Tversky in \cite{kahneman1979prospect}) offers a descriptive theory of how people actually make decisions showing that humans perceive gains, losses, and probabilities in a skewed (nonlinear) manner, typically overweighting low probabilities and underweighting high probabilities.  While a large literature on prospect theory exists in economics, relatively little research has investigated such behavioral decision-making by defenders and/or attackers, and its effects  on CPS security (exceptions include \cite{9030279,7544460,sanjab2017prospect,abdallah2019impacts,9069456,abdallah2020morshed}). 

These exceptions have focused on the impact of probability weighting on a single defender's decisions via decision-theoretic analysis (with no strategic attacker)~\cite{9030279}, on multiple defenders' investments in networks (with the emphasis being on understanding the role of the network structure)~\cite{7544460,abdallah2019impacts,9069456,abdallah2020morshed}, 
or on behavioral decision-making by both players for settings with a single target  \cite{sanjab2017prospect}. In contrast to these works, we consider the effects of behavioral decision-making in a setting with multiple targets with different values to the players (i.e., the defender and the attacker). Such heterogeneity in the values of the targets comes from the nature of the CPS and level of importance to the  defender~\cite{humayed2017cyber}.

In this paper, we introduce prospect theory into a game-theoretic framework involving an attacker and a defender. Specifically, we consider a CPS consisting of many assets, and assume that the defender misperceives the probabilities of successful compromise of each asset. 
We first establish the convexity of the objective function of each player (i.e., attacker and defender), and we use this to prove the existence of a pure strategy Nash equilibrium (PNE) for the Behavioral Multi-Target Security Game. 
We then show the uniqueness of that PNE in our game. We then characterize the optimal investment strategies by the (rational) players. We then show that the defender and the attacker invest more in higher value assets (under appropriate conditions). Subsequently, we show via numerical simulations that nonlinear perceptions of probability can induce defenders to shift more of their investments to the more valuable assets, thereby potentially increasing their (true) expected loss. 
\section{The Multi-Target Security Game Framework}\label{sec:background}

In this section, we introduce the defender model, the adversary model, and the players' utilities. 

\subsection{Strategic Defender}
Let $ \mathcal{D} $ be a defender who is responsible for defending a set $ V = \{v_1,v_2,\dots,v_n\} $ of assets. For each compromised asset $ v_{m} \in V$, defender $ \mathcal{D} $ will incur a financial loss $L_{m} \in \mathbb{R_{> \texttt{0}}}$. To reduce the attack success probabilities on assets, the defender can allocate security resources on these assets, subject to the constraints described below.

Let $n = |{V}|$.  We assume that defender $\mathcal{D}$ has a security budget $B \in \mathbb{R}_{\ge 0}$.  Thus, we define the defense strategy space of the defender by
\begin{equation}
X \triangleq \lbrace{\mathbf{x} \in \mathbb{R}^{n}_{\geq \texttt{0}}: \!\!\sum_{v_{i} \in V} \!\!x_{i} \leq B  \rbrace}.
\label{eq:defense_strategy_space}
\end{equation}
In other words, the defense strategy space for defender $\mathcal{D}$ consists of all non-negative investments on assets such that the sum of all investments does not exceed the budget $B$. 
We denote any particular vector of investments by defender $\mathcal{D}$ by $\mathbf{x} \in X$.

\subsection{Strategic Attacker}
Let $ \mathcal{A} $ be an attacker who is attempting to compromise the set $ V $ of assets.\footnote{It is realistic to assume that multiple targets could be attacked at one time. Therefore, we allow the attacker to launch simultaneous attacks on different multiple targets.} 
For each compromised asset $ v_{m} \in V$, the attacker $ \mathcal{A} $ will incur a financial gain $G_{m} \in \mathbb{R_{> \texttt{0}}}$. To increase the attack success probabilities on assets, the attacker can allocate attack resources on these assets, subject to a budget constraint $P \in \mathbb{R_{\geq \texttt{0}}}$. Thus, we define the attack strategy space of the attacker by
\begin{equation}
Y \triangleq \lbrace{\mathbf{y} \in \mathbb{R}^{n}_{\geq \texttt{0}} : \!\!\sum_{v_{i} \in V} \!\!y_{i} \leq P  \rbrace}.
\label{eq:attack_strategy_space}
\end{equation}
In other words, the attack strategy space for attacker $\mathcal{A}$ consists of all non-negative attack investments on assets, with the sum of all these investments not exceeding $P$. We denote the attacker's investment vector by $\mathbf{y} \in Y$. 

\subsection{Defender's and Attacker's Utilities}
The investments made by the defender and the attacker on each asset changes the probability that the asset can be successfully compromised by the attacker. Specifically, let $p_{i}:\mathbb{R}^{2}_{\geq 0}\rightarrow [0,1]$ be a function mapping the total defense investment $x_{i}$ and the total attack investment $y_i$ on the asset $v_i$ to an attack success probability.

The goal of defender $ \mathcal{D} $ is to choose her investment vector $\mathbf{x}$ in order to best protect her assets from being attacked. Mathematically, this is captured via the cost function
\begin{equation}\label{eq: defender rational cost node}
\overline{C}_{\mathcal{D}}(\mathbf{x},\mathbf{y}) =  \sum_{v_{i} \in V } L_{i} \hspace{1mm} p_{i}(x_{i},y_{i})
\end{equation}
subject to $\mathbf{x} \in X$. In particular, for any given $\mathbf{y} \in Y$, defender $\mathcal{D}$ chooses her investment $\mathbf{x} \in X$ to minimize $\overline{C}_{\mathcal{D}}(\mathbf{x},\mathbf{y})$.

The goal of the attacker $ \mathcal{A} $ is to choose her attack investment vector $\mathbf{y}$ in order to compromise her target assets. Mathematically, this is captured via the utility function
\begin{equation}\label{eq: attacker rational utility node}
\overline{U}_{\mathcal{A}}(\mathbf{x},\mathbf{y}) =  \sum_{v_{i} \in V } G_{i} \hspace{1mm} p_{i}(x_{i},y_{i})
\end{equation}
subject to $\mathbf{y} \in Y$. For any given $\mathbf{x} \in X$, attacker $\mathcal{A}$ chooses $\mathbf{y} \in Y$ to maximize $\overline{U}_{\mathcal{A}}(\mathbf{x},\mathbf{y})$.

Note that $ \overline{C}_{\mathcal{D}}(\mathbf{x},\mathbf{y}) $ and $\overline{U}_{\mathcal{A}}(\mathbf{x},\mathbf{y})$ are functions of both the defense investments $\mathbf{x}$ of the defender and the attack investments $\mathbf{y}$ by the attacker. 

The recent work \cite{guan2017modeling} studies this setting and provides a method to calculate the optimal investments (with respect to the cost \eqref{eq: defender rational cost node} and utility \eqref{eq: attacker rational utility node} functions, respectively). However, as mentioned in the introduction, humans have been shown to systematically misperceive probabilities, which can impact the decisions that defenders and attackers make in the presence of risk.  In the next section, we will review certain classes of probability weighting functions that capture this phenomenon, and then subsequently introduce such functions into the above Multi-Target Security Game formulation.

\section{The Behavioral Multi-Target Security Game}\label{sec:behavioralclasses}

In this section, we incorporate behavioral biases into the two player simultaneous move game formulation between the defender $\mathcal{D}$, and the attacker $\mathcal{A}$.

\subsection{Nonlinear Probability Weighting}

\begin{figure}
\begin{center}
  \includegraphics[width=0.7\linewidth]{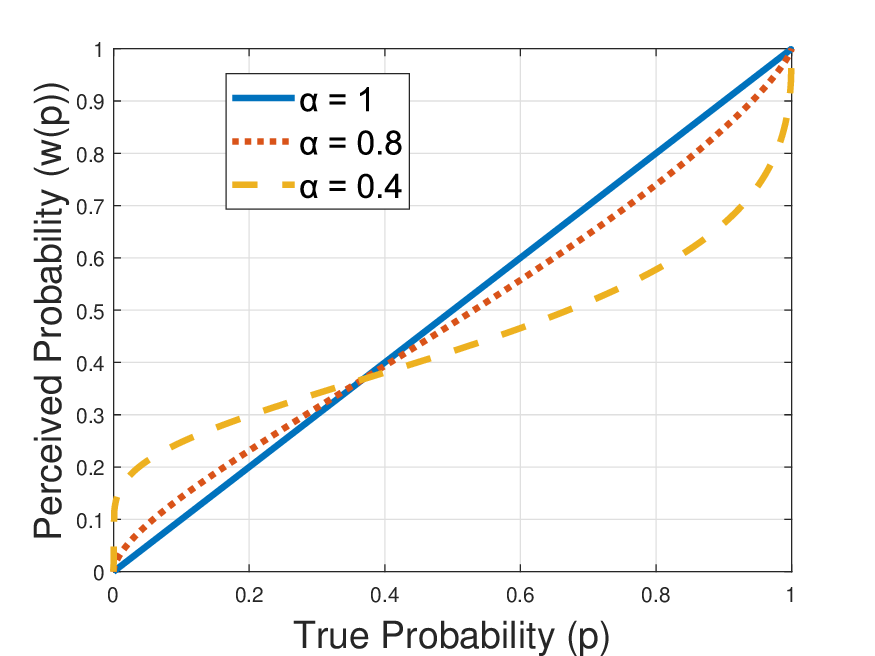}
  \caption{Prelec Probability weighting function which transforms true probabilities $ p $  into perceived probabilities $ w(p) $. The parameter $\alpha$ controls the extent of overweighting and underweighting.}
  \label{fig:Prelec Probability weighting function}
\end{center}
\end{figure}

The behavioral economics and psychology literature has shown that humans consistently misperceive probabilities by overweighting low probabilities and underweighting high probabilities \cite{kahneman1979prospect,prelec1998probability}.  More specifically, humans perceive a ``true'' probability $p \in [0,1]$ as $w(p) \in [0,1]$, where $w(\cdot)$ is a probability weighting function.  A commonly studied probability weighting function was proposed by Prelec in \cite{prelec1998probability}, and is given by
\begin{equation}\label{eq:prelec}
w(p) = \exp\Big[-(-\log(p)\hspace{0.2mm})^{\alpha}\hspace{0.5mm}\Big] ,  \hspace{3mm} p\in [0,1],
\end{equation}
where $\alpha \in (0,1]$ is a parameter that controls the extent of overweighting and underweighting.  When $\alpha = 1$, we have $w(p) = p$ for all $p \in [0,1]$, which corresponds to the situation where probabilities are perceived correctly.  Smaller values of $\alpha$ lead to a greater amount of overweighting and underweighting, as illustrated in 
Fig. \ref{fig:Prelec Probability weighting function}.  

Recall that the defender seeks to protect a set of assets, while the attacker is seeking to compromise them. The probability of each asset being successfully compromised is itself determined by the corresponding investments on that asset by both the attacker and the defender. This motivates a broad class of games that incorporate probability weighting, as defined below.

\subsection{Behavioral Multi-Target Security Game Formulation}
\begin{definition}\label{def:2}
We define a {\it Behavioral Multi-Target Security Game} as a game between an attacker and a defender for a set of targets, where both defender and attacker misperceive the attack probability on each asset according to the probability weighting function defined in \eqref{eq:prelec}. Specifically, the perceived attack probability on an asset $ v_{i} \in V$ by player $k \in \{\mathcal{A},\mathcal{D}\}$ is given by: 
\begin{equation*}
w_{k}( p_{i}(x_{i},y_{i})) = \exp\Big[-(-\log(p_{i}(x_{i},y_{i}))\hspace{0.2mm})^{\alpha_{k}}\hspace{0.5mm} \Big], \hspace{0.2mm} 
\end{equation*}
where $p_{i}(x_i,y_i)\in [0,1]$, $\alpha_k \in (0,1].$
\end{definition}
\begin{remark}
The subscript $k$ in $\alpha_{k}$ and $w_{k}(\cdot)$ allows each player (i.e., attacker and defender) in the Behavioral Multi-Target Security Game to have a different level of mis-perception.
However, for ease of notation, we will drop the subscript $k$ for most of our analysis, when it is clear from the context (i.e., any result for the defender has $\alpha = \alpha_{\mathcal{D}}$ and any result for the attacker has $\alpha = \alpha_{\mathcal{A}}$).
\end{remark}

Now, we present the optimization problem perceived by the behavioral defender and the behavioral attacker, respectively. 
\subsubsection{Defender Cost Function Minimization Problem}
\begin{equation}\label{eq: defender cost node}
\begin{aligned}
& \underset{\mathbf{x} \in X}{\text{minimize}}
& & C_{\mathcal{D}}(\mathbf{x},\mathbf{y}) =  \sum_{i=1}^{n} L_{i} \enspace w_{\mathcal{D}}(p_{i}(x_{i},y_{i})).
\end{aligned}
\end{equation}

\subsubsection{Attacker Utility Function Maximization Problem}
\begin{equation}\label{eq: attacker utility node}
\begin{aligned}
& \underset{\mathbf{y} \in Y}{\text{maximize}}
& & U_{\mathcal{A}}(\mathbf{x},\mathbf{y}) =  \sum_{i=1}^{n} G_{i} \enspace w_{\mathcal{A}}(p_{i}(x_{i},y_{i})).
\end{aligned}
\end{equation}

In a Behavioral Multi-Target Security Game, a collection of best response strategies ($\mathbf{x^{*}} ,\mathbf{y^{*}}$) is a Pure-strategy Nash Equilibrium (PNE) if and only if both equations \eqref{eq: defender nash equil simultaneous} and \eqref{eq: attacker nash equil simultaneous} below are satisfied simultaneously:
\begin{equation} \label{eq: defender nash equil simultaneous}
\mathbf{x^{*}} \in  \underset{\mathbf{x} \in X}{\text{argmin}} \enspace C_{\mathcal{D}}(\mathbf{x},\mathbf{y^{*}})
\end{equation}
\begin{equation}\label{eq: attacker nash equil simultaneous}
\mathbf{y^{*}} \in \underset{\mathbf{y} \in Y}{\text{argmax}} \enspace U_{\mathcal{A}}(\mathbf{x^{*}},\mathbf{y}).
\end{equation}

We will start by proving the existence of a PNE in the Behavioral Multi-Target Security Game, and then subsequently characterize properties of the investments by the players.  In particular, we focus on the simultaneous move game in this paper (e.g., as considered in  \cite{guan2017modeling} and the literature on Colonel Blotto games \cite{roberson2006colonel}), and leave a similar investigation of sequential games for future work.\footnote{The recent work~\cite{9304311} studied such sequential game, however with only two assets. Studying the general case with many assets would be of interest.}

\section{Existence of Pure Strategy Nash Equilibrium}\label{sec: PNE_existence}
In this section, we  prove 
the existence of a PNE for the Behavioral Multi-Target Security Game defined in Section \ref{sec:behavioralclasses}. Throughout, let the function $p_{i}({x_i,y_i})$ represent the true probability of successful attack on an asset $v_i \in V$ when the total defense and attack investments on that asset are $x_i$ and $y_i$, respectively. We make the following assumption on $p_{i}(x_i,y_i)$.

\begin{assumption} \label{ass: prob_attack}
The probability of successful attack on asset $v_i \in V$, $p_{i}(x_i,y_i)$, has the following properties.
\begin{itemize}
\item $p_{i}(x_i,y_i)$ is twice differentiable with $p_{i}(x_{i},0)= 0 $ and $\displaystyle \lim_{x_{i}\to\infty} p(x_{i},y_{i})=0 \enspace \forall y_i \in \mathbb{R_{\geq \texttt{0}}}$.
\item $p_{i}(x_i,y_i)$ is decreasing and log-convex\footnote{This is a common assumption in the literature~\cite{baryshnikov2012security,gordon2002economics}.} in $x_i$.
\item $p_{i}(x_i,y_i)$ is increasing and concave in $y_i$.
\item $\displaystyle p_{i}(x_i,y_i) \frac{\partial^2 p(x_i,y_i)}{\partial x_i \partial y_i} \leq \frac{\partial p_i(x_i,y_i)}{\partial x_i} \frac{\partial p_i(x_i,y_i)}{\partial y_i}$.
\end{itemize}
\end{assumption}

In other words, the larger the defensive security investment on a target, the less
likely that the target will be successfully attacked. On the other hand, the larger the attack resources used to attack a target, the higher the chance that the target is compromised successfully. The assumptions of concavity and twice-differentiability are common in literature~\cite{guan2017modeling, 9030279}. 

A particular success function which we will focus on throughout this work is
\begin{equation} \label{eq: defend-attack expon probability}
p_{i}(x_i,y_i)= \exp(- x_{i} - a_{i}) (1 - \exp(-y_{i})),
\end{equation}
where $a_{i} \in \mathbb{R}_{\ge 0}$ in \eqref{eq: defend-attack expon probability} represents the pre-existing (or inherent) security investments on a node, which decrease the successful attack probability even under no additional defense investment. Such probability functions fall within the class commonly considered in security economics  \cite{gordon2002economics,baryshnikov2012security}, and satisfy the conditions in Assumption~\ref{ass: prob_attack}.

\begin{lemma} \label{lemma: perceived convex in defender investment}
For every asset $v_{i} \in V$, the perceived probability of attack $w(p_{i} (x_{i},y_{i}))$ is convex in the defense investment $x_i$ under Assumption \ref{ass: prob_attack}.
\end{lemma}
\begin{proof}
For ease of notation, we drop the subscript $i$ in the following analysis.
First note that since $0 \leq p (x,y)\leq 1$, we have $0 \leq -\log(p(x,y)) \leq \infty $ for all $x$ and $y$. Substituting $p(x,y)$ into the probability weighting function defined in \eqref{eq:prelec}, we have
$$
 w(p(x,y)) = \exp\Big[ -(-\log(p(x,y)))^\alpha\Big].
$$
Now, calculating the second partial derivative of $w(p(x,y))$ w.r.t $x$ yields
\begin{flalign*}
\frac{\partial w(p(x,y))}{\partial x} &= \alpha w(p(x,y))
(-\log(p(x,y)))^{\alpha-1} \frac{\frac{\partial p(x,y)}{\partial x}}{p(x,y)},
\end{flalign*}
\begin{flalign*}
\pdv[2]{w(p(x,y))}{x}  &=  \alpha \frac{w(p(x,y)) (-\log(p(x,y)))^{\alpha-1} }{(p(x,y))^2} \\ 
& \left[ (1-\alpha) \left(\frac{\partial p(x,y)}{\partial x}\right)^2 (-\log(p(x,y)))^{-1} \right. \\
& + \alpha \left(\frac{\partial p(x,y)}{\partial x}\right)^2 (-\log(p(x,y)))^{\alpha-1} \\ 
& + \left. \left( p(x,y)\pdv[2]{p(x,y)}{x} - \left(\frac{\partial p(x,y)}{\partial x}\right)^2 \right) \right].
\end{flalign*}
Since $0 < \alpha \leq 1$, the first term on the R.H.S. of $\pdv[2]{w(p(x,y))}{x}$ is always non-negative. The second term is also non-negative. Also, $p(x,y)$ is twice-differentiable and log-convex in $x$, and the feasible defense strategy domain $X$ is convex. Therefore,  $p(x,y)\pdv[2]{p(x,y)}{x} \geq (\frac{\partial p(x,y)}{\partial x})^2$ \cite{boyd2004convex} which ensures that the third term is also non-negative. Therefore, $w(p_{i}(x_i,y_i))$ is convex in the defense investment $x_i$.
\end{proof}

\begin{lemma} \label{lemma: perceived concave in attacker investment}
Under Assumption \ref{ass: prob_attack}, if $ p_{i}(x_i,y_i) \in [0,\frac{1}{e}) \forall x_i, y_i \in \mathbb{R_{\geq \texttt{0}}}$, then 
\begin{enumerate}
    \item The perceived probability $w(p_{i} (x_i,y_i))$ will be concave in the attack investment $y_i$.
    \item The partial derivative $\displaystyle \frac{\partial^2 w(p(x_i,y_i))}{\partial x_i \partial y_i}$ is negative.
\end{enumerate}
\end{lemma}
\begin{proof}
(i) Beginning with $\pdv[2]{w(p(x,y))}{y}$ (which has the same form as $\pdv[2]{w(p(x,y))}{x}$ in the proof of Lemma \ref{lemma: perceived convex in defender investment}), we have
\begin{flalign*}
\pdv[2]{w(p(x,y))}{y}  &=  \alpha \frac{w(p(x,y)) (-\log(p(x,y)))^{\alpha-1} }{(p(x,y))^2} \\ 
& \bigg[ (1-\alpha) \left(\frac{\partial p(x,y)}{\partial y}\right)^2 (-\log(p(x,y)))^{-1}  \\
& + \alpha \left(\frac{\partial p(x,y)}{\partial y}\right)^2 (-\log(p(x,y)))^{\alpha-1} \\ 
& + \left( p(x,y)\pdv[2]{p(x,y)}{y} - \left(\frac{\partial p(x,y)}{\partial y}\right)^2\right) \bigg].
\end{flalign*}
Since $p(x,y) < \frac{1}{e}$, $(-\log(p(x,y)))^{-1} < 1$ and $(-\log(p(x,y)))^{\alpha-1} < 1$. Moreover, $\alpha (-\log(p(x,y)))^{\alpha-1} + (1-\alpha) (-\log(p(x,y)))^{-1} < 1$. Therefore, the summation of the first, second and fourth term is negative. From Assumption \ref{ass: prob_attack}, $\pdv[2]{p(x,y)}{y} < 0$ which implies that the third term is also negative. Therefore, $\pdv[2]{w(p(x,y))}{y} < 0$, i.e.,  $w(p_{i}(x_i,y_i))$ is concave in the attacker investment $y_i$.

(ii) The proof of (ii) is similar to that of part (i) and Lemma~\ref{lemma: perceived convex in defender investment} by using the second partial derivative formula and Assumption~\ref{ass: prob_attack}, and thus we omit its steps.
\end{proof}

Note that for attack success probabilities given by \eqref{eq: defend-attack expon probability}, the condition $p_i(x_i, y_i) \in [0, \frac{1}{e})$ is guaranteed when the inherent defenses of the asset (given by parameter $a_i$) satisfy $a_i \ge 1$.

This brings us to the following result, establishing the existence of a PNE in the Behavioral Multi-Target Security Games.

\begin{theorem}
Under Assumption \ref{ass: prob_attack}, if $p_i(x_i,y_i) \in [0,\frac{1}{e}) \forall x_i, y_i \in \mathbb{R_{\geq \texttt{0}}}$, a PNE exists in the Behavioral Multi-Target Security Game.
\end{theorem}
\begin{proof}
From \eqref{eq:defense_strategy_space} and \eqref{eq:attack_strategy_space}, the strategy spaces $X$ and $Y$ are compact and convex. Let the Hessian matrices of $C_{\mathcal{D}}(\mathbf{x},\mathbf{y})$ (in \eqref{eq: defender cost node}) and $U_{\mathcal{A}}(\mathbf{x},\mathbf{y})$ (in \eqref{eq: attacker utility node}) be $H_\mathcal{D}$ and $H_\mathcal{A}$, respectively. Both $H_\mathcal{D}$ and $H_\mathcal{A}$ are diagonal by definition since $p_i(x_i,y_i)$ for each asset only depends on $x_i$ and $y_i$. Moreover, from Lemma \ref{lemma: perceived convex in defender investment}, each diagonal element in $H_\mathcal{D}$ is non-negative and therefore $C_{\mathcal{D}}(\mathbf{x},\mathbf{y})$ is continuous and convex in $\mathbf{x}$. Similarly, Lemma \ref{lemma: perceived concave in attacker investment} shows that each diagonal element in $H_\mathcal{A}$ is non-positive and thus 
$U_{\mathcal{A}}(\mathbf{x},\mathbf{y})$ is continuous and  concave in $ \mathbf{y} $. Therefore, a pure-strategy Nash equilibrium exists for our Behavioral Multi-Target Security Game~\cite{glicksberg1952further,rosen1965existence}.
\end{proof}

After establishing the existence of a PNE in our  Behavioral Multi-Target Security Game, we study the characteristics of the investments of the players (the defender and the attacker) in the game.

\section{Properties of the Optimal Investment Decisions}\label{sec:properties_at_PNE}
In this section, we characterize properties of the optimal investment decisions by the players. 

\subsection{Uniqueness of PNE}
We first show the uniqueness of the PNE for the Behavioral Multi-Target Security Game (defined in Section~\ref{sec:behavioralclasses}).

\begin{theorem}\label{thm:uniqueness_pne}
Suppose that the asset values for the defender and attacker share a common ordering (i.e., $L_{1} \geq L_{2} \geq  \dots \geq  L_n$ and  $G_{1} \geq G_{2} \geq  \dots \geq  G_n$). Under Assumption~\ref{ass: prob_attack},  if $ p_{i}(x_i,y_i) \in [0,\frac{1}{e}) \forall x_i, y_i \in \mathbb{R_{\geq \texttt{0}}}$, then the PNE of the Behavioral Multi-Target Security Game is unique.
\end{theorem}
\begin{proof}
To prove the uniqueness of the PNE, we follow the argument of Rosen~\cite{rosen1965existence} by proving that the weighted non-negative sum of our payoff functions is diagonally strictly concave.

Let us denote the payoff functions of the defender and attacker as $\phi_1(\mathbf{x},\mathbf{y})$ and $\phi_2(\mathbf{x},\mathbf{y})$, respectively. Note that  $\phi_1(\mathbf{x}) = - C_\mathcal{D}(\mathbf{x},\mathbf{y})$ and $\phi_2(\mathbf{x}) =  U_\mathcal{A}(\mathbf{x},\mathbf{y})$. Now, define $\mathbf{r} = [r_1 \enspace r_2]$, and let us define $\sigma(\mathbf{x},\mathbf{y}, \mathbf{r})$ as the weighted non-negative sum of the two payoff functions $\phi_1(\mathbf{x},\mathbf{y})$ and $\phi_2(\mathbf{x},\mathbf{y})$ as follows:
\begin{multline*}
    \sigma(\mathbf{x},\mathbf{y}, \mathbf{r}) = \sum_{i=1}^{2} r_i \hspace{1mm} \phi_i(\mathbf{x},\mathbf{y}) \\
     = -r_1 \sum_{i=1}^{n} L_i \hspace{1mm} w_{\mathcal{D}}(p_{i}(x_{i},y_{i})) + r_2 \sum_{i=1}^{n} G_i \hspace{1mm} w_{\mathcal{A}}(p_{i}(x_{i},y_{i})).
\end{multline*}

Now, let us define the function $g(\mathbf{x},\mathbf{y}, \mathbf{r})$ as follows:
\begin{align*}
    g(\mathbf{x},\mathbf{y}, \mathbf{r}) &= \begin{bmatrix}
    r_1 \hspace{1mm} \nabla_{\mathbf{x}} \hspace{1mm} \phi_1(\mathbf{x},\mathbf{y})  \\
    r_2 \hspace{1mm} \nabla_{\mathbf{y}} \hspace{1mm} \phi_2(\mathbf{x},\mathbf{y})
    \end{bmatrix} = \begin{bmatrix}
    -r_1 \hspace{1mm} \nabla_{\mathbf{x}} \hspace{1mm} C_\mathcal{D}(\mathbf{x},\mathbf{y})  \\
    r_2 \hspace{1mm} \nabla_{\mathbf{y}} \hspace{1mm} U_\mathcal{A}(\mathbf{x},\mathbf{y})
    \end{bmatrix} \\ &= \begin{bmatrix}
    -r_1 \hspace{1mm} L_1 \hspace{1mm} \frac{\partial (w_{\mathcal{D}}(p_1(x_1,y_1)))}{\partial x_1}  \\
    -r_1 \hspace{1mm} L_2 \hspace{1mm} \frac{\partial (w_{\mathcal{D}}(p_2(x_2,y_2)))}{\partial x_2}  \\ \vdots \\
    -r_1 \hspace{1mm} L_n \frac{\partial (w_{\mathcal{D}}(p_n(x_n,y_n)))}{\partial x_n} \\
    r_2 \hspace{1mm} G_1 \hspace{1mm} \frac{\partial (w_{\mathcal{A}}(p_1(x_1,y_1)))}{\partial y_1}  \\ \vdots \\
    r_2 \hspace{1mm} G_n \hspace{1mm} \frac{\partial (w_{\mathcal{A}}(p_n(x_n,y_n)))}{\partial y_n}
    \end{bmatrix}.
\end{align*}

To show that $\sigma(\mathbf{x},\mathbf{y}, \mathbf{r})$ is diagonally strictly concave, it is sufficient to show that the symmetric matrix $[G(\mathbf{x},\mathbf{y},\mathbf{r}) + G^{T}(\mathbf{x},\mathbf{y},\mathbf{r})]$ is negative definite for some $\mathbf{r} > \mathbf{0}$ where $\mathbf{x} \in \mathbb{R}^{n}$ and $\mathbf{y} \in \mathbb{R}^{n}$, where $G(\mathbf{x},\mathbf{y},\mathbf{r})$ is the Jacobian with respect to $\mathbf{x}$ and $\mathbf{y}$ of $g(\mathbf{x},\mathbf{y}, \mathbf{r})$~\cite{rosen1965existence}. 

Now, we can write $G(\mathbf{x},\mathbf{y},\mathbf{r})$ as 
\begin{align*}
    G(\mathbf{x},\mathbf{y},\mathbf{r}) &= \begin{bmatrix}
    G_1(\mathbf{x},\mathbf{y},\mathbf{r}) & G_2(\mathbf{x},\mathbf{y},\mathbf{r})  \\
     G_3(\mathbf{x},\mathbf{y},\mathbf{r}) & G_4(\mathbf{x},\mathbf{y},\mathbf{r})
    \end{bmatrix},
\end{align*}
where $G_1(\mathbf{x},\mathbf{y},\mathbf{r})$, $G_2(\mathbf{x},\mathbf{y},\mathbf{r})$, $G_3(\mathbf{x},\mathbf{y},\mathbf{r})$, and $G_4(\mathbf{x},\mathbf{y},\mathbf{r})$ each have dimension $n \times n$ and are given by:
\begin{align*}
 \scalemath{0.8}{
G_1(\mathbf{x},\mathbf{y},\mathbf{r}) = r_1 \diag{ -L_1 \pdv[2]{w_{\mathcal{D}}(p(x_1,y_1))}{x_1}, \cdots, -L_n \pdv[2]{w_\mathcal{D}(p(x_n,y_n))}{x_n}},
}
\end{align*}
\begin{align*}
 \scalemath{0.8}{
G_2(\mathbf{x},\mathbf{y},\mathbf{r}) = r_1 \diag{  -L_1 \frac{\partial^2 w_\mathcal{D}(p(x_1,y_1))}{\partial x_1 \partial y_1}, \cdots, -L_n \frac{\partial^2 w_\mathcal{D}(p(x_n,y_n))}{\partial x_n \partial y_n}},
}
\end{align*}
\begin{align*}
 \scalemath{0.83}{
G_3(\mathbf{x},\mathbf{y},\mathbf{r}) = r_2 \diag{  G_1 \frac{\partial^2 w_\mathcal{A}(p(x_1,y_1))}{\partial y_1 \partial x_1}, \cdots,G_n \frac{\partial^2 w_\mathcal{A}(p(x_n,y_n))}{\partial y_n \partial x_n}},    
}
\end{align*}    
\begin{align*}
 \scalemath{0.83}{
G_4(\mathbf{x},\mathbf{y},\mathbf{r}) = r_2 \diag{  G_1 \pdv[2]{w_\mathcal{A}(p(x_1,y_1))}{y_1}, \cdots,G_n \pdv[2]{w_\mathcal{A}(p(x_n,y_n))}{y_n}}.   
}
\end{align*}   

Now, define the symmetric real matrix $M(\mathbf{x},\mathbf{y},\mathbf{r})$ as
$$
M(\mathbf{x},\mathbf{y},\mathbf{r}) =  [G(\mathbf{x},\mathbf{y},\mathbf{r}) + G^{T}(\mathbf{x},\mathbf{y},\mathbf{r})].
$$

Now, we prove that $M(\mathbf{x},\mathbf{y},\mathbf{r})$ is negative definite by showing that $\mathbf{u}^{T} M(\mathbf{x},\mathbf{y},\mathbf{r}) \mathbf{u} < 0$ for all non-zero vectors $\mathbf{u} = 
\begin{bmatrix}
    u_{1}  &
    u_{2} &
    \dots &
    u_{2n} \end{bmatrix}^\intercal$ as follows:
\begin{multline}\label{eq:neg_def_uniqueness_pne}
\mathbf{u}^{T} M(\mathbf{x},\mathbf{y},\mathbf{r}) \mathbf{u} =  -2r_1 \left(\sum_{i=1}^{n} u^2_i L_i \pdv[2]{w_\mathcal{D}(p(x_i,y_i))}{x_i}\right) \\ + 2 r_2 \left(\sum_{i=1}^{n} u^2_{n+i} G_i \pdv[2]{w_\mathcal{A}(p(x_i,y_i))}{y_i}\right) \\ + 2  \sum_{i=1}^{n} u_i u_{n+i}  \left(-r_1 L_i  \frac{\partial^2 w_{\mathcal{D}}(p(x_i,y_i))}{\partial x_i \partial y_i} \right. \\ \left. + r_2 G_i \frac{\partial^2 w_{\mathcal{A}}(p(x_i,y_i))}{\partial y_i \partial x_i} \right).  
\end{multline}
In \eqref{eq:neg_def_uniqueness_pne}, we have $\pdv[2]{w_{\mathcal{D}}(p(x_i,y_i))}{x_i} > 0 \forall i=1,\dots,n$  (since $p_i(x_i,y_i) \in [0,\frac{1}{e})$, it follows directly from the proof of Lemma~\ref{lemma: perceived convex in defender investment}), $L_i > 0$ (from defender's financial loss definition), and $u^2_i \geq 0$. Moreover, since  $\pdv[2]{w_{\mathcal{A}}(p(x_i,y_i))}{y_i} < 0 \forall i=1,\dots,n$  (from Lemma~\ref{lemma: perceived concave in attacker investment}), $G_i > 0$ (from attacker's financial gain definition), the summation of the first and second term is always negative. Moreover, from Lemma~\ref{lemma: perceived concave in attacker investment}(ii), we have $\frac{\partial^2 w_{\mathcal{D}}(p(x_i,y_i))}{\partial x_i \partial y_i} < 0$ and $\frac{\partial^2 w_{\mathcal{A}}(p(x_i,y_i))}{\partial y_i \partial x_i} < 0$. Thus, choosing 
$$
r_1 = \frac{1}{L_1 \abs{\left.\frac{\partial^2 w_{\mathcal{D}}(p(x_i,y_i))}{\partial x_i \partial y_i}\right|_{(x_i^{*},y_i^{*}) \in \argmin_{x_i,y_i} \frac{\partial^2 w_{\mathcal{D}}(p(x_i,y_i))}{\partial x_i \partial y_i} }}},
$$
$$
r_2 = \frac{1}{G_n \abs{\left.\frac{\partial^2 w_{\mathcal{A}}(p(x_i,y_i))}{\partial y_i \partial x_i}\right|_{(\bar{x}_i,\bar{y}_i) \in \argmax_{x_i,y_i} \frac{\partial^2 w_{\mathcal{A}}(p(x_i,y_i))}{\partial y_i \partial x_i} }}}
$$
where $(x_i^*,y_i^*)$ denote the investments on asset $v_i$ with minimum $\frac{\partial^2 w_{\mathcal{D}}(p(x_i,y_i))}{\partial x_i \partial y_i}$ across the $n$ assets and $(\bar{x}_i,\bar{y}_i)$ denote the investments on asset $v_i$ with maximum $\frac{\partial^2 w_{\mathcal{A}}(p(x_i,y_i))}{\partial y_i \partial x_i}$ across the $n$ assets. Note that this choice minimizes $r_1$ by choosing the maximum possible value of its denominator since $\frac{\partial^2 w_{\mathcal{D}}(p(x_i,y_i))}{\partial x_i \partial y_i} < 0$. Similarly, this choice maximizes $r_2$ by choosing the minimum possible value of its denominator since $\frac{\partial^2 w_{\mathcal{A}}(p(x_i,y_i))}{\partial y_i \partial x_i} < 0$. Therefore, this ensures that the third term is non-positive.  Therefore, we have $\mathbf{u}^{T} M(\mathbf{x},\mathbf{y},\mathbf{r}) \mathbf{u} < 0$ and thus $\sigma(\mathbf{x},\mathbf{y},\mathbf{r})$ is diagonally strictly concave for some $\mathbf{r} > \mathbf{0}$.

From Theorem 2 in \cite{rosen1965existence}, since $\sigma(\mathbf{x},\mathbf{y},\mathbf{r})$ is diagonally strictly concave for some $\mathbf{r} > \mathbf{0}$, the equilibrium point of the Behavioral Multi-Target Security Game is unique.
\end{proof}

\subsection{Locations of Optimal Investments}
We next characterize the optimal investments by the defender for a given set of investments by the attacker, and then do the same for the attacker. In particular, we denote the optimal investments by $\mathbf{x^{*}}(\alpha_{\mathcal{D}})$ and $\mathbf{y^{*}}(\alpha_{\mathcal{A}})$ to indicate that such investments will depend on the probability weighting parameters $\alpha_{\mathcal{D}}$ and $\alpha_{\mathcal{A}}$, respectively.

\begin{proposition}\label{prop: optimal defense allocation ordered}
Consider a defender $\mathcal{D}$. Let the true probability of successful attack on each asset be given by \eqref{eq: defend-attack expon probability}. Consider a set of $n$ assets whose losses can be put in the descending order  $L_{1} \geq L_{2} \geq  \dots \geq  L_n$. Suppose $y_1 \geq y_2 \geq \dots \geq y_n \ge 0$, and that the pre-existing defense investments on each asset satisfy $a_1 = a_2 = \dots = a_n$. Then, the optimal defense allocation of \eqref{eq: defender cost node}, denoted  $\mathbf{x^{*}}(\alpha_{\mathcal{D}})=
\begin{bmatrix}
    x^{*}_{1}(\alpha_{\mathcal{D}})  &
    x^{*}_{2}(\alpha_{\mathcal{D}}) &
    \dots &
    x^{*}_{n}(\alpha_{\mathcal{D}}) \end{bmatrix}^\intercal
$, has the property that  $x^{*}_{1}(\alpha_{\mathcal{D}}) \geq x^{*}_{2}(\alpha_{\mathcal{D}}) \geq \dots\geq x^{*}_{n}(\alpha_{\mathcal{D}})$.
\end{proposition}

\begin{proof}
From the KKT conditions for the defender's best response, for every pair of nodes $i$ and $j$ with nonzero optimal investments by the defender, the marginals satisfy 
$$
L_i \frac{\partial (w_{\mathcal{D}}(p_i(x_i,y_i)))}{\partial x_i}|_{x_i = x^{*}_i} = L_j \frac{\partial (w_{\mathcal{D}}(p_j(x_j,y_j)))}{\partial x_j}|_{x_j = x^{*}_j}. 
$$

If the probability of successful attack on the asset $v_i$ is given by \eqref{eq: defend-attack expon probability}, then using the Prelec probability weighting function \eqref{eq:prelec}, the defender's perceived probability of successful attack on $v_i$ would be
\begin{equation*}
w_{\mathcal{D}}(p_{i}(x_i,y_i)) = \exp\left(-(x_{i}+a_{i}-\log(1-e^{-y_{i}}))^{\alpha_{\mathcal{D}}}\right).
\label{eq:perceived_expon_attack}
\end{equation*}

Denoting $k_i = a_i -\log(1-e^{-y_i})$, the above marginals under the defender's best response would satisfy
\begin{equation}
L_i (x_i^* + k_i)^{\alpha_{\mathcal{D}} - 1} e^{-(x_i^* + k_i)^{\alpha_{\mathcal{D}}}} =  L_j (x_j^* + k_j)^{\alpha_{\mathcal{D}} - 1} e^{-(x_j^* + k_j)^{\alpha_{\mathcal{D}}}}
\label{eq:marginals_equal_defender}
\end{equation}
for all nodes $v_i,v_j$ with nonzero optimal investments $x_i^*$ and $x_j^*$, respectively. Now, if $y_i \geq y_j$, we have 
\begin{align*}
y_i \geq y_j 
\iff & 1 - e^{-y_i} \geq 1 - e^{-y_j}\\
\iff &  -\log(1 - e^{-y_i}) \leq  - \log(1 - e^{-y_j}) \\
\iff &  a_i -\log(1 - e^{-y_i}) \leq  a_j - \log(1 - e^{-y_j}) \\
\iff & k_i  \leq k_j,
\end{align*}
where we used the assumption that $a_i = a_j \enspace \forall i\neq j$.  

Using \eqref{eq:marginals_equal_defender} and assuming without loss of generality that $i < j$, we obtain
\begin{align*}
     L_i (x_i^* + k_i)^{\alpha_{\mathcal{D}} - 1} e^{-(x_i^* + k_i)^{\alpha_{\mathcal{D}}}} &= L_j (x_j^* + k_j)^{\alpha_{\mathcal{D}} - 1} e^{-(x_j^* + k_j)^{\alpha_{\mathcal{D}}}} \\
     \Rightarrow \frac{e^{-(x_i^* + k_i)^{\alpha_{\mathcal{D}}}}}{(x_i^* + k_i)^{1-\alpha_{\mathcal{D}}}} &= \frac{L_j}{L_i}\frac{e^{-(x_j^* + k_j)^{\alpha_{\mathcal{D}}}}}{(x_j^* + k_j)^{1-\alpha_{\mathcal{D}}}} \\
     &< \frac{e^{-(x_j^* + k_j)^{\alpha_{\mathcal{D}}}}}{(x_j^* + k_j)^{1-\alpha_{\mathcal{D}}}}
\end{align*}
since $L_i > L_j$. Note that $\frac{e^{-r^{\alpha_{\mathcal{D}}}}}{r^{1-\alpha_{\mathcal{D}}}}$ is a decreasing function of $r \in (0, \infty)$.  Thus, from the above expression, we have
\begin{align*}
x_i^* + k_i &> x_j^* + k_j \\
\Rightarrow x_i^* &= x_j^* + k_j - k_i \ge x_j^*,
\end{align*}
since $k_i \le k_j$.  This concludes the proof.
\end{proof}

The above result showed that the defender will invest more in higher-valued assets if the attacker has invested more in higher valued assets. We now show that a non-behavioral attacker will indeed prefer to invest more in higher-valued assets (even if the defender has invested more on those assets) under certain conditions, namely when there are significant differences in the values of the assets to the attacker.

\begin{proposition}\label{prop: optimal attack allocation ordered}
Consider a non-behavioral attacker $\mathcal{A}$ (i.e., $\alpha_{\mathcal{A}} = 1$) and a non-behavioral defender $\mathcal{D}$ (i.e., $\alpha_{\mathcal{D}} = 1$). Let the true probability of successful attack on each asset be given by \eqref{eq: defend-attack expon probability}. Consider a set of $n$ assets whose gains can be put in descending order $G_{1} \geq G_{2} \geq  \dots \geq  G_n$ such that $\frac{G_i}{G_j} \geq \frac{L_i}{L_j} \enspace \forall i < j$. Suppose that the pre-existing defense investments on each asset satisfy $a_1 = a_2 = \dots = a_n$. Then, 
\begin{enumerate}
    \item The attacker's investment at the PNE is given by $y_i^* = y_j^* + \log(\frac{G_i}{G_j}) -  \log(\frac{L_i}{L_j}) \forall i,j \in \{1,\dots,k_{\mathcal{A}}\}$ where $k_{\mathcal{A}}$ is the number of nodes that have nonzero attack investment at PNE. Formally, $k_{\mathcal{A}}$ is the largest $k$ such that  $P - \log(\frac{\prod_{i=1}^{k} G_i}{G_k^{k}}) +  \log(\frac{\prod_{i=1}^{k} L_i}{L_k^{k}}) > 0$.

    \item The defender's investment at the PNE is given by $x_i^* = x_j^* + \log(\frac{L_i}{L_j}) \forall i,j \in \{1,\dots,k_{\mathcal{D}}\}$ where $k_{\mathcal{D}}$ is the number of nodes that have nonzero defense investment at PNE. Formally, $k_{\mathcal{D}}$ is the largest $k$ such that  $B - \log(\frac{\prod_{i=1}^{k} L_i}{L_k^{k}}) > 0$.
\end{enumerate}
\end{proposition}

\begin{proof}
From \eqref{eq: defender nash equil simultaneous} and \eqref{eq: attacker nash equil simultaneous}, we prove the PNE investments by showing that the defender's PNE investment is the defender's best response to the attacker's PNE investment and that the attacker's PNE investment is the attacker's best response to the defender's PNE investment.

(i) From the KKT conditions for the attacker's best response, for every pair of nodes $i$ and $j$ with nonzero optimal investments by the attacker, the marginals must satisfy 
$G_i \frac{\partial (p_i(x_i^*,y_i))}{\partial y_i}|_{y_i = y^{*}_i} = G_j \frac{\partial (p_j(x_j^*,y_j))}{\partial y_j}|_{y_j = y^{*}_j}$.

For the probability function \eqref{eq: defend-attack expon probability}, this condition becomes
$$
G_i e^{-x^*_i-a_i} e^{-y_i^*}  = G_j e^{-x^*_j-a_j} e^{-y_j^*}    
$$
for all nodes $i,j$ with nonzero optimal investments $y^{*}_i$ and $y^{*}_j$, respectively. Taking the logarithm of both sides and rearranging, we have 
\begin{equation}\label{eq:inv_att_rational}
y_i^* = y_j^* + \log\left(\frac{G_i}{G_j}\right)- x_i^* + x_j^*,
\end{equation}
where we used the assumption that $a_i = a_j \enspace \forall i \neq j$.

Now, substituting with the defender's  
investment $x_i^* = x_j^* + \log(\frac{L_i}{L_j})$ in \eqref{eq:inv_att_rational} yields 
\begin{equation}\label{eq:inv_attacker_rational}
y_i^* = y_j^* + \log\left(\frac{G_i}{G_j}\right) - \log\left(\frac{L_i}{L_j}\right).
\end{equation}
This shows that (i) is the attacker's best response to (ii).

Now, we derive the attack PNE investment on each node. First, note from \eqref{eq:inv_attacker_rational} that if an asset $v_j$ has nonzero attack investment at the PNE, since $\frac{G_i}{G_j} \geq \frac{L_i}{L_j} \forall i < j$, all assets $v_i$ with $i < j$ would have also nonzero attack investment at the PNE as well. Formally, we have $y_1^* \geq y_2^* \geq \dots \geq y_n^*$.

Suppose that the PNE investments are such that only the top $k$ nodes ($v_1, v_2,\dots, v_k)$ get nonzero investments from the attacker, and the remaining nodes ($v_{k+1},\dots, v_n)$ get zero investment. Substituting the PNE attack investments of all assets $y_2^*,\dots,y_k^*$  in terms of the PNE attack investment of the first asset $y_1^*$ from \eqref{eq:inv_attacker_rational} into the budget constraint $\sum_{i=1}^{k} y_i^* = P$ yields
\begin{align*}
 y_1^* + \underset{i \neq 1}{\sum_{i=2}^{k}} \left( y_1^* + \log\left(\frac{G_i}{G_1}\right)- \log\left(\frac{L_i}{L_1}\right) \right)  = P \\
\implies  k y_1^* + \underset{i \neq 1}{\sum_{i=2}^{k}} \log\left(\frac{G_i}{G_1}\right) - \underset{i \neq 1}{\sum_{i=2}^{k}} \log\left(\frac{L_i}{L_1}\right)  = P \\
\implies   k y_1^* + \log(\frac{\prod_{i=2}^{k} G_i}{G_1^{k-1}}) - \log(\frac{\prod_{i=2}^{k} L_i}{L_1^{k-1}})  = P \\
\implies  y_1^* = \frac{P - \log(\frac{\prod_{i=1}^{k} G_i}{G_1^{k}}) +  \log(\frac{\prod_{i=1}^{k} L_i}{L_1^{k}})}{k}.
\end{align*}
Thus, the PNE attack investment on the remaining assets is calculated by substituting the derived $y_1^*$ in \eqref{eq:inv_attacker_rational} which yields
$$
y_i^* = \frac{P - \log(\frac{\prod_{i=1}^{k} G_i}{G_i^{k}}) +  \log(\frac{\prod_{i=1}^{k} L_i}{L_i^{k}})}{k}, \forall i \in \{2,\dots,k\}.
$$
To have nonzero investment on all assets $v_1,\dots,v_k$, we must have 
$$
P - \log(\frac{\prod_{i=1}^{k} G_i}{G_i^{k}}) +  \log(\frac{\prod_{i=1}^{k} L_i}{L_i^{k}}) > 0 \forall i \in \{1,\dots,k\}.
$$
However, since $y_1^* \geq y_2^* \geq \dots \geq y_k^*$, it is sufficient to have  
$$
P - \log(\frac{\prod_{i=1}^{k} G_i}{G_k^{k}}) +  \log(\frac{\prod_{i=1}^{k} L_i}{L_k^{k}}) > 0.
$$
Thus, the number of nodes that have nonzero attack investment at PNE, denoted by $k_{\mathcal{A}}$, is the largest $k$ such that the above inequality holds.

(ii) From part (i), since the number of nodes that have nonzero attack investment at PNE is $k_{\mathcal{A}}$, substituting \eqref{eq:inv_att_rational} in budget constraint $\sum_{i=1}^{k_{\mathcal{A}}}{y_i^*} = P$ yields
\begin{align*}
 y_j^* + \underset{i \neq j}{\sum_{i=1}^{k_{\mathcal{A}}}} \left( y_j^* + \log\left(\frac{G_i}{G_j}\right)- x_i + x_j \right)  = P \\
\implies  k_{\mathcal{A}} \hspace{1mm} y_j^* + \underset{i \neq j}{\sum_{i=1}^{k_{\mathcal{A}}}} \log\left(\frac{G_i}{G_j}\right)- \underset{i \neq j}{\sum_{i=1}^{k_{\mathcal{A}}}} x_i + \underset{i \neq j}{\sum_{i=1}^{k_{\mathcal{A}}}} x_j   = P \\
\stackrel{(a)}\implies  k_{\mathcal{A}} \hspace{1mm} y_j^* + \log(\frac{\prod_{i=1}^{k_{\mathcal{A}}} G_i}{G_j^{k_{\mathcal{A}}}}) - B + k_{\mathcal{A}} \hspace{1mm} x_j  = P \\
\implies  y_j^* = \frac{P + B -\log(\frac{\prod_{i=1}^{k_{\mathcal{A}}} G_i}{G_j^{k_{\mathcal{A}}}})}{k_{\mathcal{A}}} - x_j
\end{align*}
for any node $v_j \in \{v_1,\dots,v_{k_\mathcal{A}} \}$. Note that (a) holds since $y_i^* = 0 \forall i > k_{\mathcal{A}}$. Thus, from Assumption~\ref{ass: prob_attack}, we have $p_i(x_i,0) = 0$ and thus at the PNE, we have $x_i  = 0 \forall i > k_{\mathcal{A}}$. 

Now, substituting $y^*_j$ in the defender's cost \eqref{eq: defender rational cost node} yields 
\begin{align*}
\overline{C}_{\mathcal{D}}(\mathbf{x},\mathbf{y}^*) &= \sum_{i=1}^{k_{\mathcal{A}}}  L_i e^{-x_i-a_i} (1 - e^{-y_i^*}) \\
&= \sum_{i=1}^{k_{\mathcal{A}}} L_i \left(e^{-x_i-a_i} - e^{-a_i - \frac{P + B - \log(\frac{\prod_{j=1}^{{k_{\mathcal{A}}}} G_j}{G_i^{{k_{\mathcal{A}}}}})}{{k_{\mathcal{A}}}}} \right). 
\end{align*}

Now, from the KKT conditions for the defender's best response, every pair of nodes $i$ and $j$ with nonzero optimal investments, the marginals must satisfy  $$
\frac{\partial (\overline{C}_{\mathcal{D}}(x,y^*))}{\partial x_i}|_{x_i = x^{*}_i} = \frac{\partial (\overline{C}_{\mathcal{D}}(x,y^*))}{\partial x_j}|_{x_j = x^{*}_j}. 
$$
Thus, we have 
$$
L_i e^{-x_i^*-a_i} = L_j e^{-x_j^*-a_j}
$$
for all nodes $i,j$ with nonzero optimal investments $x^{*}_i$ and $x^{*}_j$, respectively. Taking the logarithm of both sides and rearranging, we have 
\begin{equation}\label{eq:inv_def_rational}
x_i^* = x_j^* + \log\left(\frac{L_i}{L_j}\right),
\end{equation}
where we used the assumption that $a_i = a_j \enspace \forall i \neq j$. 
This shows that (ii) is the defender's best response to (i).

Similar to part (i), suppose that only the top $k'$ nodes get nonzero investments from the defender at the PNE. Substituting the PNE defense investments of all assets $x_2^*,\dots,x_{k'}^*$  in terms of the PNE defense investment of the first asset $x_1^*$ from \eqref{eq:inv_def_rational} into the budget constraint $\sum_{i=1}^{k'} x_i^* = B$ yields
$$
x_i^* = \frac{B - \log\left(\frac{\prod_{i=1}^{k'} L_i}{L_i^{k'}}\right)}{k'}, \forall i \in \{1,\dots,k'\}.
$$
Thus, the number of nodes that have nonzero defense
investment at PNE, denoted by $k_\mathcal{D}$, is the largest $k'$ such that the above inequality holds.

From the above analysis, we show that (i) and (ii) satisfies \eqref{eq: attacker nash equil simultaneous} and \eqref{eq: defender nash equil simultaneous} simultaneously and thus (i) and (ii) are the PNE investments of the attacker and the defender, respectively.
\end{proof}

The above results shows that if the asset values for the defender and attacker share a common ordering and if the values to the attacker are significantly different between the assets, then, in the PNE, both players invest more in their higher valued assets (noting that the attacker would invest the same in all assets if the ratio of gains are exactly the ratio of losses for any two assets within the CPS, i.e., $\frac{G_i}{G_j} = \frac{L_i}{L_j} \forall i < j$).  

We will show an example of such a PNE later (emphasizing the CPS defender's investments and the attacker's efforts) in our numerical simulations in Section~\ref{sec: numer_sim}.
\begin{remark}
Note that if the asset values to the attacker are the same (i.e., $G_1 = G_2 = .... = G_n$) and if $L_i \geq L_j \hspace{1mm} \forall i<j$, we have $y_i^* \leq y_j^* \hspace{1mm} \forall i<j$. This indicates that under homogeneous valuations, the non-behavioral attacker would invest more in the assets that are less-important to the defender since they are expected to be less-protected.
\end{remark}

\section{Numerical Simulations}\label{sec: numer_sim}

\begin{figure}
\begin{center}
  \includegraphics[width=0.6\linewidth]{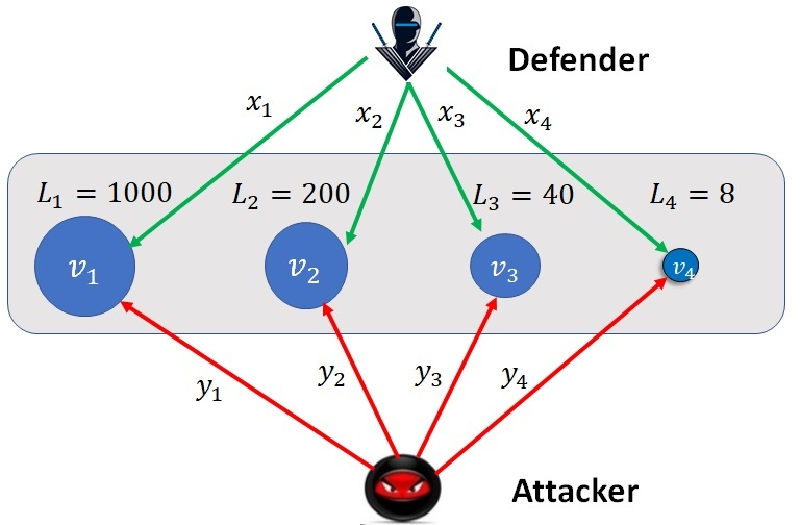}
   \caption{A simple visualization of our Multi-Target Game Setup. The green arrows are the defense resources while the red arrows are the attack efforts on the assets. The quantities $x_i$ and $y_i$ denote the amount of resources allocated to defending and attacking asset $v_i$, respectively.}
  \label{fig:Setup_Visualization}
\end{center}
\end{figure}

In this section, we provide numerical simulations results to validate our findings in Section~\ref{sec:properties_at_PNE} and to show the effect of behavioral decision-making.

\begin{figure*}[t!] 
\centering
\begin{minipage}[t]{1.0\textwidth}
\begin{minipage}[t]{.32\textwidth}
\centering
 \includegraphics[width=0.8\linewidth]{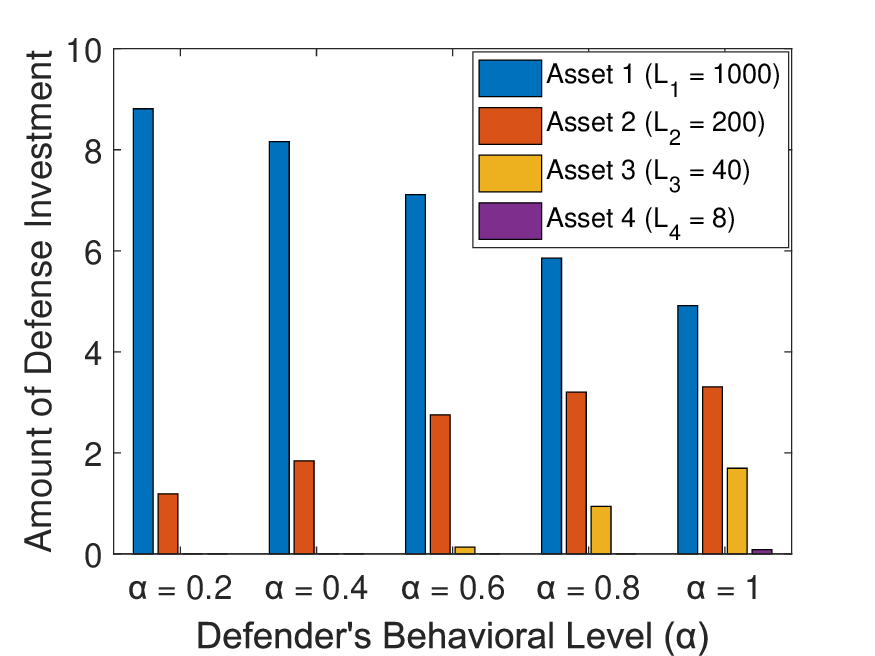}
  \caption{Effect of behavioral probability weighting on the defense investments. The asset with the highest financial loss takes higher portion of the defense investments as the defender becomes more behavioral (i.e., $\alpha$ decreases) while the attacker is non-behavioral.}
  \label{fig:Behavioral_Effect_Defense}
 \end{minipage}\hfill
\begin{minipage}[t]{.32\textwidth}
\centering
  \includegraphics[width=0.8\linewidth]{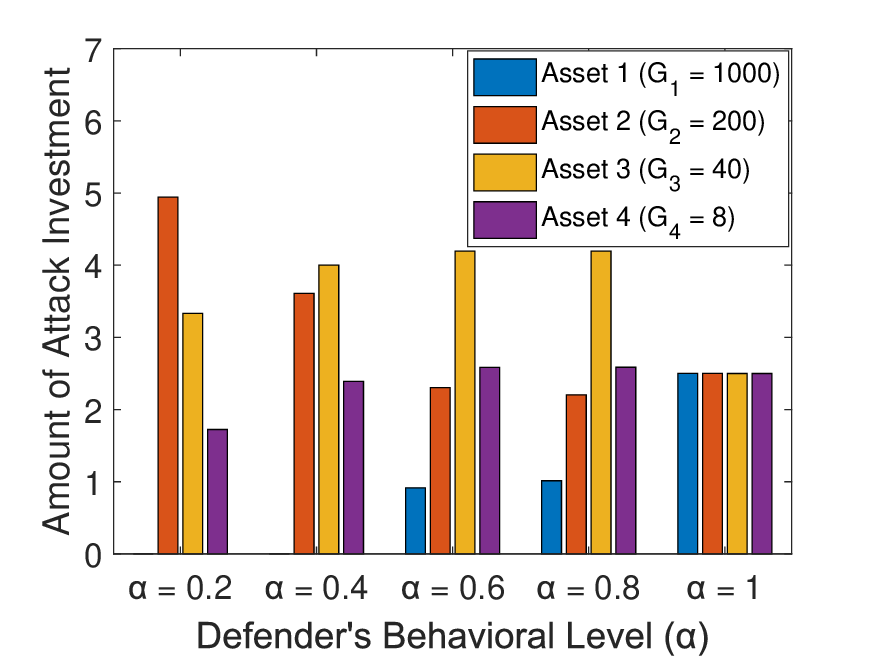}
  \caption{Effect of defender's behavioral probability weighting on  the attack investments. The asset with the highest financial gain takes much lower portion of the attack investments as the defender becomes more behavioral  while the attacker is non-behavioral.}
  \label{fig:Behavioral_Effect_Attack}
\end{minipage}\hfill
\begin{minipage}[t]{.32\textwidth}
\centering
  \includegraphics[width=0.8\linewidth]{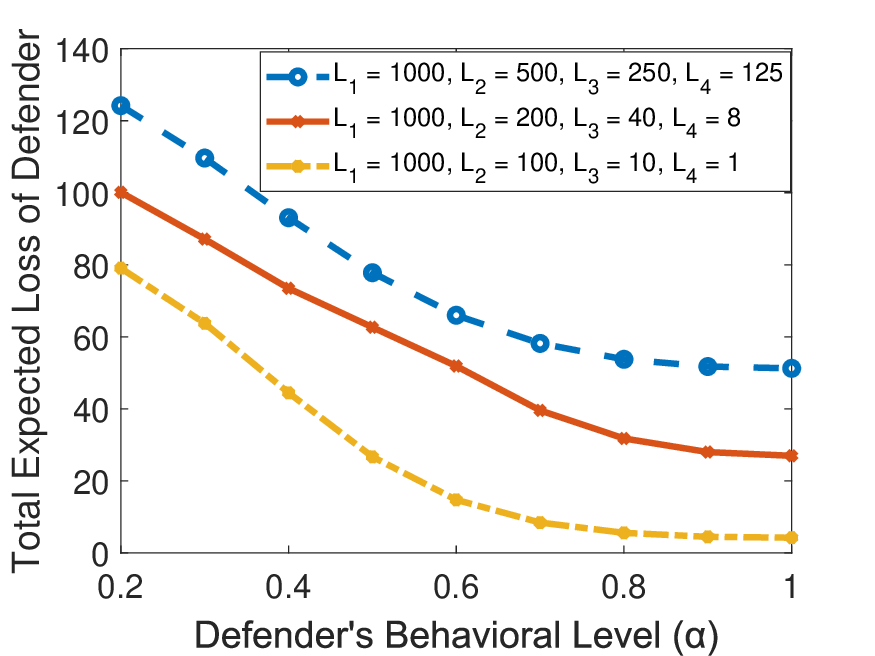}
  \caption{Effect of behavioral probability weighting on the true expected loss of the defender for different assets' loss values. The cost of the defender is worse if the defender becomes more behavioral while attacker is non-behavioral.}
  \label{fig:Behavioral_Effect_objectives}
\end{minipage}
\end{minipage}
\end{figure*}

\subsection{Experimental Setup}
We emulated four critical assets (or  targets). For the defender, the first asset has very high loss (i.e., $L_{1} = 1000$) while the second and third assets have lower losses (with $L_{2} = 200 $, $L_{3} = 40$) and the fourth asset has the least loss ($L_{4} = 8$). For the attacker, we employ symmetric gains for successful attack (i.e., $G_{1} = 1000$, $G_{2} = 200$, $G_{3} = 40$, and $G_{4} = 8$). We let the total defense budget  for defending the three critical assets and the total attack budget to compromise them be $B = 10 $ and $P = 10$, respectively. The probability of successful attack on each of the assets is given by 
$$
p(x,y) = e^{-x-1}(1-e^{-y})
$$
where $x$ and $y$ are the defense and attack investment on that asset, respectively. The above function satisfies the conditions in Assumption~\ref{ass: prob_attack}. We followed the best response dynamics notion to calculate the optimal investments of each player at the PNE. All of these optimal investments were calculated using Matlab Optimization toolbox.

\subsection{Effect of Perception on Investments}
In this subsection, we show the effect of probability misperception on the defense and attack investment decisions in the Behavioral Multi-Target Security Game. We note the ordering of defense investments on the assets (which is consistent with Proposition~\ref{prop: optimal defense allocation ordered}). Fig. \ref{fig:Behavioral_Effect_Defense} shows the difference in the defense investments for each of the assets as $\alpha_{\mathcal{D}}$ changes for the defender while keeping the attacker non-behavioral (with $\alpha_{\mathcal{A}} = 1$).  We observe that the asset with the highest financial loss takes a higher portion of the defense investments as the defender becomes more behavioral (i.e., $\alpha_{\mathcal{D}}$ decreases). Fig. \ref{fig:Behavioral_Effect_Attack} illustrates the effect of defender's behavioral level on attacker's investment decision. The non-behavioral attacker's investments facing a non-behavioral defender is consistent with Proposition~\ref{prop: optimal attack allocation ordered}. Note also that when both players are non-behavioral, the PNE investments satisfy the  condition for number of nodes with non-zero investments in Proposition~\ref{prop: optimal attack allocation ordered} (Here, we have $k_{\mathcal{D}} = k_{\mathcal{A}} = 4$). We also observe that a non-behavioral attacker would put less resources on the first asset, with the highest gain, when facing behavioral defender who ``over-protects'' this asset. The insight here that the attacker would not waste attack resources on the highly-defended asset (Asset 1) but it tries to attack the remaining assets. 
\subsection{Effect of Behavioral Investments on CPS Defender's Loss}
It is also worth considering the total expected system loss $E_T$ of the defender in equilibrium, given by the sum of the real losses of all assets.
First, we consider our previously considered loss valuations (i.e., $L_1 = 1000$, $L_2 = 200$, $L_3 = 40$, and $L_4 = 8$). As shown in Fig.  \ref{fig:Behavioral_Effect_objectives}, when the defender is non-behavioral (i.e., $\alpha = 1$) $ E_{T}  =  26.96$, while $ E_{T} =  100.12 $ when $\alpha = 0.2$ with a non-behavioral attacker in both scenarios. This considerable increase in the total real loss of the behavioral defender shows that probability weighting induces defender to invest in a sub-optimal manner, when some assets are much more valuable to the defender. Moreover, as the behavioral level increases, the effect of suboptimal investments is more pronounced in terms of the defender's total expected (true) loss.  Fig. \ref{fig:Behavioral_Effect_objectives} also shows such insight for two alternative loss valuations scenarios.

\section{Conclusion}\label{sec: conclusion}
This paper presented a game-theoretic framework that takes account of behavioral attitudes of defender and attacker in Multi-Target Security Game where the attacker and the defender place their investments to compromise and protect the target assets respectively. Specifically, we considered the scenario where the (human) defender misperceives the probabilities of successful attack in each asset. We then established the existence and uniqueness of PNE for our Behavioral Multi-Target Security Game. We then provided the optimal solutions for non-behavioral players for that game. Finally, we provided numerical simulations that validated our results and showed that nonlinear perceptions of probability can induce the defender to invest more on the assets with higher losses.
Future avenues of research would be studying the setup of a behavioral attacker and its resulting properties, and validating our findings via human subject experiments (similar to~\cite{woods2020network} on attack graphs).

\bibliographystyle{IEEEtran}
\bibliography{sample}


\end{document}